\documentclass[12pt]{article}
\usepackage{graphicx}

\newcommand{\beq}{\begin{equation}}
\newcommand{\eeq}{\end{equation}}
\newcommand{\beqn}{\begin{eqnarray}}
\newcommand{\eeqn}{\end{eqnarray}}
\newcommand{\beqns}{\begin{eqnarray*}}
\newcommand{\eeqns}{\end{eqnarray*}}

\begin{document}
\begin{titlepage}
\begin{center}

\hfill USTC-ICTS-14-20\\
\hfill October 2014\\

\vspace{2.5cm}

{\large {\bf  Asymmetries from the interference between Cabibbo-favored and doubly-Cabibbo-suppressed $D$ decays}} \vspace*{1.0cm}\\
{  Dao-Neng Gao$^\dagger$} \vspace*{0.3cm} \\
{\it\small
Interdisciplinary Center for Theoretical Study, University of Science and Technology of China,
Hefei, Anhui 230026 China}\vspace*{1cm}
\end{center}
\begin{abstract}
\noindent
 A phenomenological analysis of $D\to K\pi$ and $D_s^+\to K K $decays including both
Cabibbo-favored and doubly-Cabibbo-suppressed modes have been presented by employing the present experimental data. SU(3) symmetry breaking effects from the decay constants and form factors have been taken into account in the analysis. Three asymmetries, $R(D^0)$, $R(D^+)$, and $R(D_s^+)$,  which are generated through interference between Cabbibo-favored and doubly-Cabibbo-suppressed decays, are estimated. Theoretical results agree well with the
current measurements.
\end{abstract}

\vfill \noindent

$^{\dagger}$ Email address: ~gaodn@ustc.edu.cn
\end{titlepage}

\section{Introduction}

Two-body hadronic $D$ decays could provide useful information for the study of the weak and strong interactions. These processes contains three types: Cabibbo-favored (CF), singly-Cabibbo-suppressed (SCS), and doubly-Cabibbo-suppressed (DCS) decays. The first two types have largely been observed experimentally, while suffering from the backgrounds of CF decays, only a few channels have been measured for the third one \cite{PDG2014}. On the other hand, as pointed out by Bigi and Yamamoto \cite{BY95} (and also in Ref. \cite{Xing1997}), DCS modes involving neutral kaons may show their existence by studying some interesting asymmetries due to interference between CF transitions (producing an $s$ quark, and thus a $\bar{K}^0$) and DCS transitions (producing an $\bar{s}$ quark, and thus a $K^0$), which, for instance, can be defined as
\beq\label{rd1}
R(D)\equiv\frac{{\cal B}(D\to K_S\pi)-{\cal B}(D\to K_L\pi)}{{\cal B}(D\to K_S\pi)+{\cal B}(D\to K_L\pi)}
\eeq
for $D\to K\pi$ decays. By explicitly setting
\beq\label{ratio1}\frac{A(D\to K^0\pi)}{A(D\to \bar{K}^0\pi)}=r e^{i\phi},\eeq
which is the ratio of DCS and CF amplitudes and $\phi$ is the strong phase between them, one can get
\beq\label{rd2}R(D)=-\frac{2 r\cos\phi}{1+r^2}, \eeq
and for the small $r$, we have $R(D)\simeq -2 r \cos\phi$. Thus, the measurement of these asymmetries may help to extract some information about the DCS processes. Experimentally, these measurements have been done by the CLEO Collaboration \cite{CLEO08} as
\beq\label{rdcleo}
R(D^0)=0.108\pm 0.025\pm 0.024,\;\;\;\; R(D^+)=0.022\pm 0.016\pm 0.018.
\eeq
Similar asymmetry for $D_s^+$ induced from the decays $D_s^+\to K^+ K^0$ and $D_s^+\to K^+\bar{K}^0$, namely,
\beq\label{rds}R(D_s^+)\equiv\frac{{\cal B}(D_s^+\to K_S K^+)-{\cal B}(D_s^+\to K_L K^+)}{{\cal B}(D_s^+\to K_S K^+)+{\cal B}(D_s^+\to K_L K^+)} \eeq
will be reported by  the BES Collaboration soon \cite{BES}.

The effective Hamiltonian relevant for CF and DCS decays can be given by \beqn\label{Hamiltonian}
{\cal H}_{\rm eff}&=&\frac{G_F}{\sqrt{2}}\left\{V_{ud}V^*_{cs}
[C_1(\bar{s}_i c_i)_{V-A}(\bar{u}_j d_j)_{V-A}+C_2 (\bar{s}_i
c_j)_{V-A}(\bar{u}_j
d_i)_{V-A}]\right.\nonumber\\&&\left.+V_{us}V^*_{cd}[C_1(\bar{d}_i
c_i)_{V-A}(\bar{u}_j s_j)_{V-A}+C_2 (\bar{d}_i
c_j)_{V-A}(\bar{u}_j s_i)_{V-A}]\right\}\nonumber\\&&+{\rm H.c.},
\eeqn where $V-A$ denotes $\gamma_\mu(1-\gamma_5)$, and the summation over repeated color indices ($i$ and $j$) is understood. The first line
in eq. (\ref{Hamiltonian}) is for CF transitions and the second line for
DCS transitions. Historically, the naive factorization approach has long been utilized
in the analysis of the hadronic $D$ decays, although there is an
obvious shortcoming that it cannot lead to the scale and scheme
independence for the final physical amplitude. On the other hand, some interesting methods, such as the QCD factorization \cite{BBNS} and pQCD \cite{KLS}, which work very well for the non-leptonic $B$ decays, cannot lead to reliable predictions for $D$ decays \cite{Gao07} for the charm quark mass is not heavy enough.

In Ref. \cite{Gao07}, we have performed a phenomenological analysis of $D\to K\pi$ decays including both CF and DCS modes based on the quark-diagrammatic approach \cite{CC86}. In order to determine all decay amplitudes of these transitions using the present experimental data, some SU(3) symmetry breaking effects have been taken into account. $R(D^0)$ and $R(D^+)$ have been calculated, which are consistent with the results reported by the CLEO Collaboration \cite{CLEO08}. The purpose of the present study is twofold. First, we will generalize the study in Ref. \cite{Gao07} to the $D_s^+\to K^0 K^+$ and $D_s^+\to \bar{K}^0 K^+$ decays, since $R(D_s^+)$ of eq. (\ref{rds}) will be measured by the BES Collaboration soon. Second, we would like to reanalyze these processes since some data have been updated after the publication of Ref. \cite{Gao07}. It is easy to see that, CF and DCS $D\to K\pi$ and $D_s^+\to K K$ decays, which are guided by eq. (\ref{Hamiltonian}), are free of penguin contributions. Studies of penguin contributions might be very interesting to understand SU(3) symmetry breaking effects and/or CP violation in SCS  $D\to\pi\pi, KK$ decays \cite{SCSpaper}. However, it has been pointed out in Ref. \cite{Gao07} that the present analysis cannot be directly extended to the case of SCS processes.

The remainder of the paper is organized as follows. In section 2, we shall discuss the amplitude decompositions of $D\to K\pi$ and $D_s^+\to KK$ decays, and some useful constraints will be obtained. In Section 3, a phenomenological analysis is carried out and asymmetries $R(D)$'s will be estimated. Our main results are summarized in Section 4.

\section{Amplitude decompositions}
In terms
of the quark-diagram topologies ${\cal T}$ (color-allowed), ${\cal
C}$ (color-suppressed), ${\cal E}$ ($W$-exchange), and ${\cal A}$
($W$-annihilation) \cite{CC86}, the decay amplitudes for $D\to K\pi$ and $D_s^+\to K^0(\bar{K}^0) K^+$ transitions can be written as
\beqn\label{topoamplitude1} A(D^0\to
K^-\pi^+)=i\frac{G_F}{\sqrt{2}}V_{ud}V^*_{cs}({\cal T}+{\cal
E}),\\
\sqrt{2} A(D^0\to \bar{K}^0
\pi^0)=i\frac{G_F}{\sqrt{2}}V_{ud}V^*_{cs}({\cal
C}-{\cal E}),\\
A(D^+\to \bar{K}^0\pi^+)=i\frac{G_F}{\sqrt{2}}V_{ud}V^*_{cs}({\cal
T}+{\cal C}),\\
A(D_s^+\to \bar{K}^0 K^+)=i\frac{G_F}{\sqrt{2}}V_{ud}V^*_{cs}({\cal C}_s+{\cal A}_s),\\
A(D^0\to K^+\pi^-)=i\frac{G_F}{\sqrt{2}}V_{us}V_{cd}^*({\cal
T}^\prime+{\cal E}^\prime),\\
\sqrt{2}A(D^0\to K^0
\pi^0)=i\frac{G_F}{\sqrt{2}}V_{us}V_{cd}^*({\cal C}^\prime-{\cal
E}^{\prime}),\\
A(D^+\to K^0\pi^+)=i\frac{G_F}{\sqrt{2}}V_{us}V_{cd}^*({\cal
C}^{\prime}+{\cal A}^\prime),\\
\sqrt{2}A(D^+\to
K^+\pi^0)=i\frac{G_F}{\sqrt{2}}V_{us}V_{cd}^*({\cal
T}^\prime-{\cal A}^\prime),\\
A(D_s^+\to K^0 K^+)=i\frac{G_F}{\sqrt{2}}V_{us}V_{cd}^*({\cal T}^\prime_s+{\cal C}^\prime_s).\eeqn  For our
notations, we have extracted the CKM matrix elements and factor
$G_F/\sqrt{2}$ from the quark-diagram amplitudes, and the prime is
added to DCS amplitudes. Using the factorization hypothesis, the quark-diagram amplitudes
${\cal T}$'s and ${\cal C}$'s appearing in above equations can be further expressed as \beqn\label{factorization1}
&&{\cal T}=f_\pi(m_D^2-m_K^2) F^{D\to
 K}_0(m_\pi^2) a_1^{\rm eff},\nonumber
 \\&& {\cal C}=f_K(m_D^2-m_\pi^2)
 F^{D\to\pi}_0(m_K^2) a_2^{\rm eff},\nonumber\\&&
{\cal T}^\prime=f_K(m_D^2-m_\pi^2) F^{D\to \pi}_0(m_K^2) a_1^{\rm
eff},\nonumber\\
&&{\cal C}^\prime=f_K(m_D^2-m_\pi^2)
 F^{D\to\pi}_0(m_K^2) a_2^{\rm eff},\nonumber\\
 &&{\cal C}_s=f_K (m_{D_s}^2-m_K^2)F_0^{D_s\to K}(m_K^2) a_2^{\rm eff},\nonumber\\
 &&{\cal C}_s^\prime=f_K (m_{D_s}^2-m_K^2)F_0^{D_s\to K}(m_K^2) a_2^{\rm eff},\nonumber\\
 &&{\cal T}_s^\prime=f_K (m_{D_s}^2-m_K^2)F_0^{D_s\to K}(m_K^2) a_1^{\rm eff},
 \eeqn
where $a_i^{\rm eff}$'s are  regarded as the effective Wilson
coefficients fixed from the data (in the naive factorization,
$a_{1,2}=C_{1,2}+C_{2,1}/N_c$), and $F^{D_{(s)}\to \pi(K)}_0(q^2)$'s are the form factors for $D_{(s)}\to \pi(K)$ transitions.
\begin{figure}[t]
\begin{center}
\includegraphics[width=14cm,height=3.0cm]{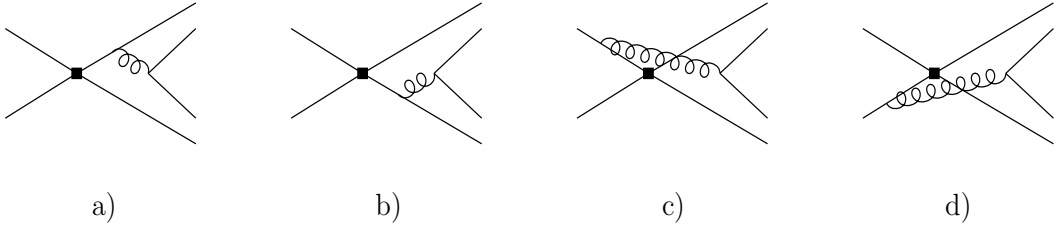}
\end{center}
\caption{$W$-exchange or
$W$-annihilation diagrams via gluon emission. The solid square denotes the
weak vertex.}
\end{figure}

For the $W$-exchange and $W$-annihilation amplitudes, it has been pointed out in \cite{Gao07, LY05} that the diagrams induced by the topologies of gluon emission
arising from the quarks of the weak vertex, as shown in Fig. 1, play important roles in the hadronic $D$
decays. This is also the case for $B$ decays \cite{BBNS01}.
This contribution has been given in Refs. \cite{LY05,BBNS01}, which reads \beqn\label{annihilation0} && {\cal E}=f_D
f_K f_\pi\frac{C_F}{N_C^2}\pi \alpha_s
C_1\left[18\left(X_A-4+\frac{\pi^2}{3}\right)+2 r_\chi^\pi
r_\chi^K X_A^2\right], \\\label{annihilation1} &&{\cal E}^\prime={\cal E},
 \\\label{annihilation2} &&{\cal A}^\prime=f_D f_K
f_\pi\frac{C_F}{N_C^2}\pi \alpha_s
C_2\left[18\left(X_A-4+\frac{\pi^2}{3}\right)+2 r_\chi^\pi
r_\chi^K X_A^2\right],\eeqn
and \beq\label{annihilations3}
{\cal A}_s=f_{D_s} f_K
f_K\frac{C_F}{N_C^2}\pi \alpha_s
C_2\left[18\left(X_A-4+\frac{\pi^2}{3}\right)+2 r_\chi^K
r_\chi^K X_A^2\right], \eeq
where $X_A$ is introduced to parameterize the logarithmically  divergent integrals due
to the end-point singularity, $C_1$, $C_2$ are the Wilson
coefficients in (\ref{Hamiltonian}), and
\beq\label{rpi}
r_\chi^P=\frac{2m^2_P}{m_c(m_1+m_2)}\eeq
with $m_{1,2}$ are the current quark mass inside the $P$ meson.  As shown in Ref. \cite{Pich98}, in the isospin limit, there exists
\beq\label{rpi1}\frac{m_\pi^2}{m_u+m_d}=\frac{m_{K^\pm}^2}{m_u+m_s}.
\eeq
This means $r_\chi^\pi=r_\chi^K$. Consequently, one can get some constraints for weak annihilation amplitudes
\beq\label{annihilation3} {\cal A}^\prime=\frac{C_2}{C_1}{\cal
E}^\prime =\frac{C_2}{C_1}{\cal
E},\eeq
and \beq\label{annihilation4}
{\cal A}_s=\frac{f_{D_s} f_K}{f_{D}f_\pi} {\cal A}^\prime.
\eeq

 Thus, from eq. (\ref{factorization1}) together with eqs. (\ref{annihilation3}) and (\ref{annihilation4}), one will find that only three complex amplitudes, chosen, for example, as ${\cal T}$, ${\cal C}$, and ${\cal E}$, are independent, which could be determined using the present experimental data. On the other hand, it is easy to see that eq. (\ref{annihilation3}) seems to be not very physical since $C_1$ and $C_2$ are both scale and scheme dependent \cite{BBL}. As shown in \cite{Gao07}, the ratio $C_2/C_1$ is about $-0.5\sim -0.3$ for the scale $\mu$ around $1.0 \sim 1.5$ GeV, which is the range of the scale relevant for $D$ decays. Eq. (\ref{annihilation3}) supports that the relative phase between ${\cal A}^\prime$ and ${\cal E}^\prime$ is $180^\circ$. Since theoretical determination for the absolute value of $C_2/C_1$ cannot be done unambiguously, in this paper, we will adopt
 \beq\label{annihilation5} {\cal A}^\prime=-\kappa~ {\cal E}^\prime=-\kappa ~{\cal E}
 \eeq
instead of eq. (\ref{annihilation3}), where $\kappa$ is the positive parameter fixed from the experimental data.

\section{Asymmetries}
In order to go into the analysis of the amplitudes from the data, first we need to know the information about the form factors $F^{D_{(s)}\to \pi(K)}_0(q^2)$. Here we shall use the same way as in Ref. \cite{Gao07}, by adopting the Bauer-Stech-Wirbel model \cite{BSW}, in which the form factors are assumed to behave as a
monopole, \beq\label{pole}F_0^{D_{(s)}\to P}(q^2)=\frac{F_0^{D_{(s)}\to
P}(0)}{1-q^2/m_*^2},\eeq where $P$ denotes $\pi$ or $K$, and $m_*$ is the pole mass, which has been shown in \cite{BSW} for $P=\pi$ or $K$.
$F_0^{D\to P}(0)$ ($P=\pi, K$) can be obtained via $F_0^{D\to P}(0)=F_+^{D\to P}(0)$, since the latter can be measured in semi-leptonic $D^0\to
\pi^- \ell^+ \nu$ and $D^0\to K^-\ell^+\nu$ decays. The latest experimental values from the CLEO Collaboration \cite{CLEO09} give \beqn && F_+^{D\to K}(0)=0.739\pm 0.007\pm 0.005,\nonumber\\ &&F_+^{D\to
\pi}(0)=0.666\pm 0.004\pm 0.003. \eeqn
Because there is no similar measurement for $F_0^{D_s\to K}$, we directly take its value from Ref. \cite{BSW} in our analysis.

\begin{table}[t]\begin{center}\begin{tabular}{c c c c}
\hline\hline\\ ${\cal T}$ [GeV$^3$]~&~${\cal C}$
[GeV$^3$]~&~${\cal E}$ [GeV$^3$]~&~
$\kappa$\\
\\$0.34\pm 0.06$~&~$(0.24\pm 0.02) e^{ \mp i(151^\circ\pm 22^\circ)}$~&~$(0.23\pm 0.12) e^{ \pm i (115^\circ\pm 19^\circ)}$~&~$0.33\pm 0.19$
\\\hline\hline\end{tabular}\caption{Numerical results of quark-diagram amplitudes
${\cal T}$, ${\cal C}$, ${\cal E}$,  and parameter $\kappa$ estimated by using the present data. }\end{center}\end{table}

Let us move to the determination of the decay amplitudes from currently available data. As mentioned in the previous section, we have three independent complex amplitudes: ${\cal T}$, ${\cal C}$, and ${\cal E}$. Without loss of generality, ${\cal T}$ is set to be real. $\delta_C$ ($\delta_E$) is the relative strong phase of ${\cal C}$ (${\cal E}$) to ${\cal T}$. Recall that the positive parameter $\kappa$ introduced in eq. (\ref{annihilation5}), totally we have six real parameters: ${\cal T}$, $|{\cal C}|$,$\delta_C$, $|{\cal E}|$, $\delta_E$, and $\kappa$,  which could be calculated from six branching ratios: ${\cal B}(D^0\to K^-\pi^+)$, ${\cal B}(D^0\to \bar{K}^0\pi^0)$, ${\cal B}(D^+\to \bar{K}^0\pi^+)$, ${\cal B}(D^0\to K^+\pi^-)$,  ${\cal B}(D^+\to K^+\pi^0)$, and ${\cal B}(D_s^+\to \bar{K}^0K^+)$, given by particle data group \cite{PDG2014}. The results of ${\cal T}$, ${\cal C}$, ${\cal E}$, and $\kappa$ are summarized in Table 1, and the error is due to the uncertainties of experimental branching ratios. Other amplitudes  such as ${\cal T}^\prime$,  ${\cal C}^\prime$,${\cal E}^\prime$, ${\cal A}^\prime$, ${\cal C}_s^{(\prime)}$, ${\cal T}_s^\prime$, and ${\cal A}_s$ can be easily derived using eqs. (\ref{factorization1}),(\ref{annihilation4}) and (\ref{annihilation5}). Note that we get $\kappa=0.33\pm 0.19$, which is consistent with the range of $C_2/C_1$ : $-0.5\sim -0.3$ used in Ref. \cite{Gao07}, and also the previous fits by the CLEO Collaboration \cite{CLEO0802} and Bhattacharya and Rosner \cite{BR08}:
\beq\label{fit08} {\cal A}^\prime=(-0.32\pm 0.24){\cal E}.
\eeq

Now we start to estimate the asymmetries $R(D)$. For the neutral $D$ decays, it has been shown in \cite{Gao07} that  \beq \label{rd00}{A(D^0\to {K}^0
\pi^0)}=-\tan^2\theta_C A(D^0\to \bar{K}^0 \pi^0),\eeq
were $\theta_C$ is the Cabibbo angle.
Consequently, one has \beq \label{rd0} R(D^0)=\frac{2
\tan^2\theta_C}{1+\tan^4 \theta_C}\simeq 2 \tan^2\theta_C\simeq 0.106, \eeq
which is in agreement with the measurement in eq. (\ref{rdcleo}). The same result has been given in Refs. \cite{BY95, JR06}. As pointed out in Ref. \cite{JR06}, the decays $D^0\to K^0\pi^0$ and $D^0\to\bar{K}^0 \pi^0$ are related to each other under the $U$-spin symmetry $s\leftrightarrow d$, thus the SU(3) symmetry breaking is expected to be extremely small in the relation (\ref{rd00}).

In the $D^+$ case, we have \beq\label{rr1}\frac{A(D^+\to K^0 \pi^+)}{A(D^+\to
\bar{K}^0 \pi^+)}=-\tan^2\theta_C \frac{{\cal C}^\prime+{\cal
A}^\prime}{{\cal C}+{\cal T}}=-\tan^2\theta_C\frac{{\cal
C}^\prime-\kappa {\cal E}}{{\cal C}+{\cal T}}. \eeq
One cannot expect a similar analytic relation  as eq. (\ref{rd00}) for neutral modes. However,  as shown above, the amplitudes ${\cal T}$, ${\cal C}$, ${\cal C}^\prime$, ${\cal E}$ and the parameter $\kappa$ appearing in eq. (\ref{rr1}) have been obtained using the present experimental data. Thus, together with eq. (\ref{rd2}), the direct numerical calculation will lead to
\beqn\label{rdplus}
R(D^+)=-0.010\pm 0.026,\eeqn
where the error is also from the uncertainties of experimental branching ratios. This result is consistent with the observed value $R(D^+)=0.022\pm 0.016\pm 0.018$ \cite{CLEO08}. Here we have corrected a sign error in the calculation of $R(D^+)$ in Ref. \cite{Gao07}, some updated experimental data for $D\to K\pi$ decays have been used, and $D_s^+\to K^0(\bar{K}^0)K^+$ decays have been included in the present analysis. $R(D^+)$ was also predicted to be $-0.006^{+0.033}_{-0.028}$ in \cite{BR08}, $-0.005\pm 0.013$ in \cite{BR10}, and $-0.019\pm 0.016$ in \cite{CC10}.

Similar work can be done for the decays $D_s^+\to K^0 K^+$ and $D_s^+\to\bar{K}^0 K^+$. Using
\beq\label{rds1}
\frac{A(D^+_s\to K^0 K^+)}{A(D_s^+\to \bar{K}^0 K^+)}=-\tan^2\theta_C\frac{{\cal T}_s^\prime+{\cal C}_s^\prime}{{\cal C}_s+{\cal A}_s}\eeq
and eqs. (\ref{factorization1}),(\ref{annihilation4}) and (\ref{annihilation5}), we obtain
\beqn\label{rdsplus}
R(D_s^+)=-0.008\pm 0.007.\eeqn
At present, there is no experimental measurement available for this asymmetry. It may be reported by the BES Collaboration soon. Theoretically, the prediction of $R(D_s^+)$ has also been given by $-0.003^{+0.019}_{-0.017}$ in \cite{BR08}, $-0.0022\pm 0.0087$ in \cite{BR10}, and $-0.008\pm 0.007$ in \cite{CC10}.

As mentioned in the Introduction, DCS modes involving neutral kaons, such as $D^0\to K^0 \pi^0$ and $D^+\to K^0\pi^+$ are suffered from the background of CF modes $D^0\to \bar{K}^0\pi^0$ and $D^+\to \bar{K}^0\pi^+$, respectively, which makes it very difficult to carry out direct measurements of these decays. It has been shown in eqs. (\ref{ratio1}) and (\ref{rd2}), that asymmetries $R(D)$'s are generated from interference between CF and DCS processes,  measurements of them will be of course helpful to extract some useful information about DCS transitions. In the $D^0$ case, the measurement of $R(D^0)$ supports the relation (\ref{rd00}), which implies
\beq\label{ratiod0}
\frac{|A(D^0\to {K}^0 \pi^0)|}{|A(D^0\to \bar{K}^0 \pi^0)|}=\tan^2\theta_C,
\eeq
and the relative strong phase between these two amplitudes vanishes.

The situation is a little different in the $D^+$ case. From our analysis, one can get
\beq\label{ratiodplus}
\frac{|A(D^+\to K^0\pi^+)|}{|A(D^+\to \bar{K}^0\pi^+)|}=\tan^2\theta_C \cdot (1.44\pm 0.09),\eeq
and the corresponding  relative strong phase $\phi=94^\circ\pm 10^\circ$ (the phase around $90^\circ$ has also been obtained in Ref. \cite{BR08}), i.e., $\cos\phi=-0.067\pm0.173$. Recall that, from eq. (\ref{rd2}), the asymmetry $R(D)$ is proportional to  $\cos\phi$, therefore the central value of $R(D^+)$ is suppressed comparing with the value of $R(D^0)$, and this also leads to the large error in our prediction (\ref{rdplus}). Although the current observation of $R(D^+)$ with large uncertainty is consistent with our results, more precise measurement of this asymmetry is encouraged in order to perform more conclusive analysis.
Similar calculation can be applied to  $D_s^+\to K K$ decays, which gives
\beq\label{ratiods}
\frac{|A(D^+_s\to {K}^0 K^+)|}{|A(D^+_s\to \bar{K}^0 K^+)|}=\tan^2\theta_C \cdot (0.70\pm 0.07),
\eeq
and the relative strong phase is $96^\circ\pm 5^\circ$. We hope forthcoming measurement of $R(D^+_s)$ by the BES Collaboration may tell us some interesting information.

\section{Concluding remarks}

We have presented a phenomenological analysis of $D\to K\pi$ and $D_s^+\to K K$ decays including both CF and DCS modes. In terms of quark-diagram approach and factorization hypothesis,  all decay amplitudes for these processes have been determined using the present data. SU(3) symmetry breaking effects from the decay constants and form factors have been taken into account in the analysis. Asymmetries $R(D)$'s due to interference between CF and DCS transitions have been evaluated, and the predictions of $R(D^0)$ and $R(D^+)$ are in agreement with the experimental data.

Comparing with Ref. \cite{Gao07},  we take the absolute value of the ratio ${\cal A}^\prime/{\cal E}$, namely $\kappa$ in this paper, as a parameter fixed from data, instead of an input. Some updated experimental $D\to K\pi$ branching ratios and the latest measurements for $F_+^{D\to P}(0)$ from the CLEO Collaboration have been used in the calculation. We also include CF and DCS $D_s^+\to K K$ decays in the present work, $R(D_s^+)$ is thus estimated, which is consistent with other predictions. It is expected that experimental measurement for $R(D_s^+)$ may come soon.

It will be interesting to extend the present formalism to describe CF and DCS $D$ decays involving $\eta$ or $\eta^\prime$ mesons. However, it is seen that the relation $r_\chi^\pi=r_\chi^K$ from eq. (\ref{rpi1}) is essential to get the constraint (\ref{annihilation4}). This relation will be complicated or explicitly violated when one includes $\eta$ or $\eta^\prime$. A further discussion of this issue is open for the future investigation.

\section*{Acknowledgements}
The author is grateful to Hai-Bo Li for helpful communications and discussions. This work was supported in part by the NSF of China under Grants No. 11075149 and 11235010.

\end{document}